\documentclass[aps,prl,twocolumn,showpacs,superscriptaddress,floatfix]{revtex4}

\usepackage{graphicx}
\usepackage{amssymb}

\begin{document}

\title{Superconductivity on the border of weak itinerant ferromagnetism in UCoGe}

\author{N. T. Huy}
\affiliation{Van der Waals - Zeeman Institute, University of
Amsterdam, Valckenierstraat~65, 1018 XE Amsterdam, The
Netherlands}
\author{A. Gasparini}
\affiliation{Van der Waals - Zeeman Institute, University of
Amsterdam, Valckenierstraat~65, 1018 XE Amsterdam, The
Netherlands}
\author{D. E. de Nijs}
\affiliation{Van der Waals - Zeeman Institute, University of
Amsterdam, Valckenierstraat~65, 1018 XE Amsterdam, The
Netherlands}
\author{Y. Huang}
\affiliation{Van der Waals - Zeeman Institute, University of
Amsterdam, Valckenierstraat~65, 1018 XE Amsterdam, The
Netherlands}
\author{J. C. P. Klaasse} \affiliation{Van der Waals -
Zeeman Institute, University of Amsterdam, Valckenierstraat~65,
1018 XE Amsterdam, The Netherlands}
\author{T. Gortenmulder} \affiliation{Van der Waals -
Zeeman Institute, University of Amsterdam, Valckenierstraat~65,
1018 XE Amsterdam, The Netherlands}
\author{A. de Visser}
\email{devisser@science.uva.nl} \affiliation{Van der Waals -
Zeeman Institute, University of Amsterdam, Valckenierstraat~65,
1018 XE Amsterdam, The Netherlands}
\author{A. Hamann}
\affiliation{Physikalisches Institut, Universit{\"a}t Karlsruhe,
D-76128 Karlsruhe, Germany}
\author{T. G{\"o}rlach}
\affiliation{Physikalisches Institut, Universit{\"a}t Karlsruhe,
D-76128 Karlsruhe, Germany}
\author{H. v. L{\"o}hneysen}
\affiliation{Physikalisches Institut, Universit{\"a}t Karlsruhe,
D-76128 Karlsruhe, Germany} \affiliation{Forschungszentrum
Karlsruhe, Institut f{\"u}r Festk{\"o}rperphysik, D-76021
Karlsruhe, Germany}

\date{\today}

\begin{abstract}
We report the coexistence of ferromagnetic order and
superconductivity in UCoGe at ambient pressure. Magnetization
measurements show that UCoGe is a weak ferromagnet with a Curie
temperature $T_{C}$= 3 K and a small ordered moment $m_{0}$= 0.03
$\mu_B$. Superconductivity is observed with a resistive transition
temperature $T_{s}$ = 0.8 K for the best sample. Thermal-expansion
and specific-heat measurements provide solid evidence for bulk
magnetism and superconductivity. The proximity to a ferromagnetic
instability, the defect sensitivity of $T_{s}$, and the absence of
Pauli limiting, suggest triplet superconductivity mediated by
critical ferromagnetic fluctuations.

\end{abstract}


\maketitle

In the standard theory for superconductivity (SC) due to Bardeen,
Schrieffer and Cooper ferromagnetic (FM) order impedes the pairing
of electrons in singlet states~\cite{Berk-PRL-1966}. It has been
argued however that on the borderline of ferromagnetism, critical
magnetic fluctuations could mediate SC by pairing the electrons in
triplet states~\cite{Fay-PRB-1980}. The discovery several years
ago of SC in the metallic ferromagnets UGe$_{2}$ (at high
pressure) ~\cite{Saxena-Nature-2000},
URhGe~\cite{Aoki-Nature-2001}, and possibly UIr (at high
pressure)~\cite{Akazawa-JPCM-2004}, has put this idea on firm
footing. However, later work provided evidence for a more
intricate scenario in which SC in UGe$_{2}$ and URhGe is driven by
a magnetic transition between two polarized
phases~\cite{Pfleiderer-PRL-2002,Sandeman-PRL-2003,Levy-Science-2005},
rather than by critical fluctuations associated with the zero
temperature transition from a paramagnetic to a FM phase. Here we
report a novel ambient-pressure FM superconductor UCoGe. Since SC
occurs right on the borderline of FM order, UCoGe may present the
first example of SC stimulated by critical fluctuations associated
with a FM quantum critical point (QCP).

UCoGe belongs to the family of intermetallic U{\it TX} compounds,
with {\it T} a transition metal and {\it X} is Si or Ge, that was
first manufactured by Tro{\'c} and Tran ~\cite{Troc-JMMM-1988}.
UCoGe crystallizes in the orthorhombic TiNiSi structure (space
group $P_{nma}$)~\cite{Lloret-PhDThesis-1988,Canepa-JALCOM-1996},
just like URhGe. From magnetization, resistivity ($T \geq 4.2$ K)
~\cite{Troc-JMMM-1988,Lloret-PhDThesis-1988} and specific heat
measurements ($T \geq 1.2$ K)~\cite{Buschow-JAP-1990} it was
concluded that UCoGe has a paramagnetic ground state. This
provided the motivation to alloy URhGe (Curie temperature $T_{C} =
9.5$ K) with Co in a search for a FM QCP in the series
URh$_{1-x}$Co$_{x}$Ge ($x \leq 0.9)$~\cite{Sakarya-JALCOM-2007}.
Magnetization data showed that $T_{C}$ upon doping first
increases, has a broad maximum near $x=0.6$ ($T_{C}^{max}$= 20 K)
and then rapidly drops to 8 K for $x=0.9$
~\cite{Sakarya-JALCOM-2007}. This hinted at a FM QCP for $x
\lesssim 1.0$. In this Letter we show that the end ($x=1.0$)
compound UCoGe is in fact a weak itinerant ferromagnet. Moreover,
metallic ferromagnetism coexists with SC below 0.8 K at ambient
pressure.

Polycrystalline UCoGe samples were prepared with nominal
compositions U$_{1.02}$CoGe (sample \#2) and
U$_{1.02}$Co$_{1.02}$Ge (sample \#3) by arc melting the
constituents (natural U 99.9\%, Co 99.9\% and Ge 99.999\%), under
a high-purity argon atmosphere in a water-cooled copper crucible.
The as-cast samples were annealed for ten days at 850 $^\circ$C.
Samples for the different experiments were cut by spark erosion,
after which the defected surface was removed by polishing. Powder
X-ray diffraction patterns at $T = 300$ K confirmed the TiNiSi
structure. The lattice constants extracted are $a = 6.845$~\AA, $b
= 4.206$~\AA ~and $c = 7.222$~\AA, in agreement with literature
~\cite{Canepa-JALCOM-1996}. The phase homogeneity of the annealed
samples was investigated by electron micro-probe analysis. The
matrix has the 1:1:1 composition and all samples contained a small
amount (2\%) of impurity phases.

The dc-magnetization was measured for temperatures $T \geq 2$ K
and magnetic fields $B \leq 5$ T in a SQUID magnetometer. The
demagnetizing factor of our samples is small ($N \approx 0.08$)
and corrections due to the demagnetizing field were neglected.
Four-point low-frequency ac-resistivity and ac-susceptibility data
were obtained using a phase-sensitive bridge in the range $T =
0.02-8$~K. The specific heat was measured using a semi-adiabatic
method employing a mechanical heat switch on a sample with mass 3
g for $T= 0.5-10 $~K and with a weak thermal link on a sample with
mass 0.1 g for $T = 0.1-1.0 $~K. Thermal expansion data were
collected using a capacitance dilatometer for $T = 0.23-8$~K.

\begin{figure}
\includegraphics[width=7.5cm]{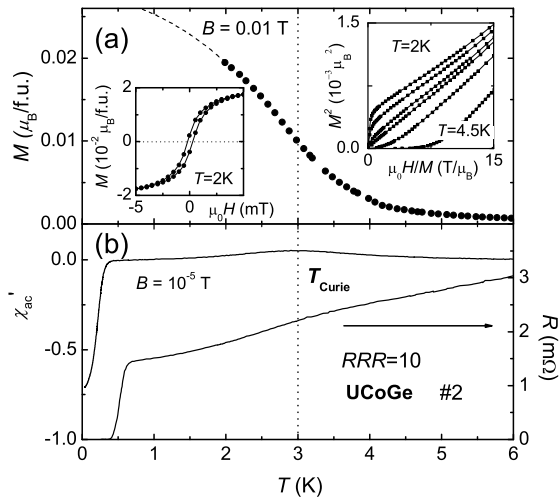}
\caption{Magnetic and SC properties of UCoGe sample \#2. (a)
Magnetization $M$ as a function of $T$ in a field $B$ of 0.01 T.
The dashed line extrapolates to $m_{0} \approx 0.03~ \mu_B$ for $T
\rightarrow 0$. The Curie temperature $T_{C} =$ 3 K is marked by
the dotted vertical line. Left inset: Hysteresis loop $M(B)$ at $T
=$ 2 K with coercive field of 0.3~mT. Right inset: Arrott plot of
magnetization isotherms at $T=$ 2.0, 2.4, 2.8, 3.0, 3.5 and 4.5 K
(from top to bottom). (b) Ac-susceptibility $\chi _{ac} ^{\prime}$
(left axis) (in $B= 10^{-5}$ T), and resistance $R$ (right axis)
as a function of $T$. The maximum in $\chi _{ac} ^{\prime}$ and
the broad hump in $R$ locate $T_{C}$. SC for sample \#2 is found
below 0.61 K in $R(T)$ and below 0.38 K in $\chi _{ac}
^{\prime}(T)$.}
\end{figure}

In Fig.~1a we show $M$ as a function of $T$ (obtained after field
cooling). The inflection point in $M(T)$ at 3 K signals a FM
transition with an unusually small ordered moment $m_{0}$. At the
lowest temperature (2 K) the transition is not complete yet, but
from the curvature of $M(T)$ the size of $m_{0}$ is estimated to
0.03 $\mu_B$. FM order is further corroborated by the hysteresis
loop in $M(B)$ at 2 K with a small coercive field of 0.3~mT (see
left inset in Fig.~1a). In the right-hand inset of Fig.~1 we show
$M$ measured at fixed $T$ between 2 and 4.5 K in an Arrott plot
(i.e. $M^{2} ~vs~ \mu_{0}H/M$). The isotherm that intersects the
origin determines the Curie temperature $T_{C}$. We extract
$T_{C}$ = 3 K, in agreement with the $M(T)$ data. The FM
transition at 3 K shows up as a broad peak in the
ac-susceptibility $\chi _{ac}^\prime (T)$ (Fig.~1b) and a hump in
the resistance $R(T)$ (Fig.~1b). The magnetic transition is a
robust property, as $M(T)$, $\chi _{ac}^\prime (T)$ and $R(T)$
data taken on samples prepared from different batches almost
coincide. The small ratio of $m_{0}$ to the effective moment
($p_{eff} = 1.7~ \mu_{B}$~\cite{Troc-JMMM-1988}) shows UCoGe is a
weak itinerant
ferromagnet~\cite{Wohlfarth-PhysicaB-1977,Moriya-Spinger-1985}.

\begin{figure}
\includegraphics[width=7.cm]{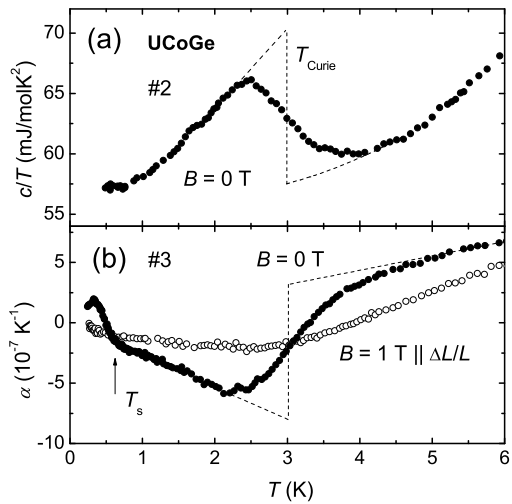}
\caption{(a) Specific heat divided by temperature, $c/T$, versus
$T$ in zero magnetic field for sample \#2. The idealized
transition (dashed line) has a step size $\Delta (c/T)$ =  0.014
J/molK$^{2}$ at $T_{C} =$ 3 K. Bulk SC for sample \#2 sets in at
0.38 K (measured by $\chi _{ac} ^{\prime}$), but the $c(T)$ data
extend down to 0.5 K only. (b) Thermal expansion coefficient
$\alpha (T)$ for sample \#3. The large negative contribution below
$\sim$ 5 K is due to FM order. The dashed line gives the idealized
transition in $\alpha (T)$ with $\Delta \alpha = -1.1$x$10^{-6}$
K$^{-1}$. The total relative length change $\Delta L/L =
(L(0.23$K$)-L(T))/L$ associated with magnetic order is obtained by
integrating $\alpha_{mag} (T)$ ({\it i.e.} the difference between
the experimental data and the linear term $\alpha = aT$ with $a =
1.1$x$10^{-7}$ K$^{-2}$ expected in the absence of FM order) and
amounts to $+1.9$x$10^{-6}$. The peak below $\sim 0.6$ K is the
thermodynamic signature of the SC transition. In a field of 1 T,
applied along the dilatation direction $\Delta L/L$, the magnetic
transition is smeared and SC is not resolved.}
\end{figure}

In Fig.~2 we show the specific heat $c(T)$ and the linear thermal
expansion coefficient, $\alpha (T) = L^{-1}dL/dT$, around the
magnetic transition. The transition width is large ($\Delta T_{C}
\sim 1$ K). The relative change $\Delta (c/T_{C})/(c/T_{C})$
assuming an ideal transition (see dashed line in Fig.~2a) is only
25 \% and the magnetic entropy associated with the transition is
small (0.3\% of $R$ln2) as expected for a weak itinerant
ferromagnet~\cite{Wohlfarth-PhysicaB-1977} with a small ordered
moment. The linear term in the electronic specific heat $\gamma$
amounts to 0.057 J/molK$^2$, which indicates UCoGe is a correlated
metal, but the electron interactions are relatively weak. In
$\alpha (T)$ the magnetic transition appears as a large negative
contribution. The size of the idealized sharp step $\Delta \alpha$
is $-1.1$x10$^{-6}$ K$^{-1}$ at $T_{C}$ = 3 K (see dashed line in
Fig.~2b) and presents a relative change $\Delta \alpha /\alpha$ of
$\approx 3.3$. This shows the magnetic transition is a bulk
phenomenon.

Below 1 K UCoGe becomes superconducting as seen by a transition to
zero in the resistance $R(T)$ and a large diamagnetic signal in
$\chi _{ac} ^{'}(T)$, see Fig.~1b and Fig.~3a. Unlike the magnetic
properties, the SC properties depend sensitively on the quality of
the samples as measured by the residual resistance ratio
$RRR=R(300$K$)/R(1$K$)$. For sample \#2 ($RRR = $ 10) and \#3
($RRR =$ 25) SC is found with resistive onset temperatures of 0.61
K (Fig.~1b) and 0.82 K (Fig.~3a), respectively. In these
polycrystalline samples the SC transition is relatively broad
($\Delta T_{s} \approx 0.15$ K). The in-phase component of the
ac-susceptibility $\chi _{ac} ^{\prime}$ starts to drop when the
resistive transition is complete. The drop is accompanied by a
small dissipative peak in the out-of-phase signal $\chi _{ac}
^{\prime \prime}$ (not plotted). At the lowest $T$ the diamagnetic
screening reaches a value of 60-70\% of the ideal screening value
$\chi_{M} =$ -1/(1-N). This indicates UCoGe is a type II SC which
is always in the mixed phase. A similar
observation~\cite{Aoki-Nature-2001} with a comparable screening
fraction was made for URhGe. Because of the intrinsic FM moments
the local field is non-zero and the magnitude of $\chi _{ac}
^{\prime}$ is reduced.

\begin{figure}
\includegraphics[width=8cm]{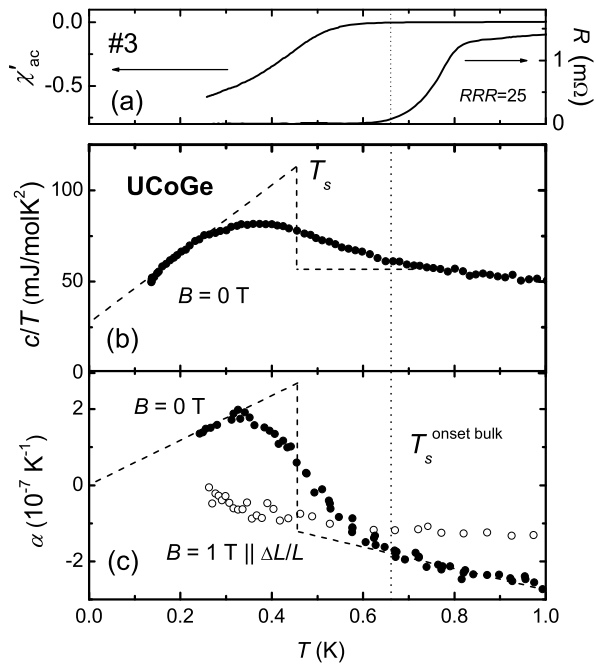}
\caption{Superconducting properties of UCoGe sample \#3. (a)
Ac-susceptibility $\chi _{ac} ^{\prime}$ (left axis) in $B =
10^{-5}$ T and resistance $R$ (right axis). (b) Specific heat
divided by temperature $c/T$ as a function of $T$. Bulk SC occurs
below $T_{s}^{onset} \approx$ 0.66 K (dotted vertical line).
Dashed line: idealized SC transition using an equal entropy
construction with a finite $\gamma$-value in the SC state of 0.028
J/molK$^{2}$. (c) Coefficient of linear thermal expansion $\alpha
(T)$. Bulk SC is observed below $T_{s}^{onset} \approx 0.66$ K.
Dashed line: idealized sharp transition with $T_{s} =$ 0.45 K. For
$B = 1$ T applied along the dilatation direction $\Delta L/L$ the
SC transition is no longer resolved.}
\end{figure}

Proof for bulk SC is obtained by specific heat (Fig.~3b) and
thermal expansion measurements (Fig.~3c). The specific heat
plotted as $c/T$ versus $T$ shows a broad transition with
$T_{s}^{onset} \approx 0.66$ K, which is almost equal to the
temperature at which the resistance becomes zero. A rough estimate
for the step-size of the idealized transition (dashed line in
Fig.~3b) in the specific heat (at $T_{s} \approx 0.45$ K) is
$\Delta (c/T_{s})/ \gamma \approx 1.0$, which is smaller than for
a conventional SC (the BCS value is 1.43), but comparable to the
value~\cite{Aoki-Nature-2001} for URhGe. In the thermal expansion
an equivalent broad SC transition is observed. Upon entering the
SC state $\alpha(T)$ shows a steady increase. We estimate the
step-size $\Delta \alpha \approx 3.8$x$10^{-7}$ K$^{-1}$, assuming
an ideal sharp transition (see dashed line in Fig.~3c) at $T_{s}
=$ 0.45 K. This step-size is comparable to the ones (with opposite
sign) extracted from thermal expansion measurements on the
heavy-fermion superconductors
URu$_2$Si$_2$~\cite{VanDijk-PRB-1995} and
UPt$_3$~\cite{VanDijk-JLTP-1993,VanDijk-PRB-1993}. In a magnetic
field of 1 T SC is suppressed and the thermodynamic signature of
the transition is no longer resolved (see Fig.~3c). The
$\alpha(T)$-data also show that magnetism and SC coexist. The
total relative length change associated with SC, obtained by
integrating $\alpha_{sc}(T)$ after correcting for the normal-state
linear contribution $\alpha = aT$ with $a = -2.7$x$10^{-7}$
K$^{-2}$ (see dashed line for 0.45 K $\leq T \leq$ 1 K in Fig.~3c)
amounts to $\Delta L/L = -0.1$x$10^{-6}$ and is small compared to
the length change $\Delta L/L = +1.9$x$10^{-6}$ due to magnetic
ordering (see caption Fig.~3). Thus magnetism is not expelled
below $T_{s}$ and coexists with SC.

\begin{figure}
\includegraphics[width=5.5cm]{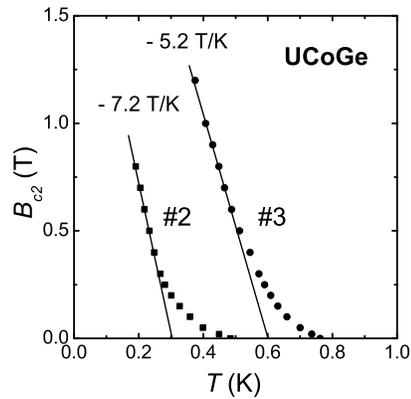}
\caption{Upper critical field $B_{c2}$ determined by the
mid-points of the resistive transitions measured in fixed magnetic
fields. The solid lines indicate $B_{c2}/dT = $ -7.2 T/K and -5.2
T/K for sample \#2 and \#3, respectively, and extrapolate to
zero-field $T_s$ values of 0.30 K and 0.60 K.}
\end{figure}

In Fig.~4 we show the upper critical field $B_{c2}(T)$ for samples
\#2 and \#3. The curvature (or tail) of $B_{c2}$ is attributed to
sample inhomogeneities. The quasi-linear behavior of $B_{c2}(T)$
at high fields extrapolates to SC transitions in zero field at
0.30 K and 0.60 K. These values are close to $T_{s}^{onset}$ for
bulk SC. From the slope $dB_{c2}/dT$ and the values of $\gamma$
and the residual resistivity $\rho_0$, we can make a crude
estimate~\cite{Orlando-PRB-1979} for the coherence length ($\xi$)
and the mean free path ($\ell$). For sample \#3 $dB_{c2}/dT =$
-5.2 T/K and $\rho_{0} = 12~ \mu \Omega$cm, and we calculate $\xi
\approx 150$~\AA ~and~ $\ell \approx 500$~\AA. This indicates
sample \#3 satisfies the clean-limit condition ($\ell > \xi$), a
prerequisite for unconventional SC~\cite{Millis-PRB-1988}. For the
less pure sample \#2 we find $\xi \approx ~200$~\AA~and~ $\ell
\approx 300$~\AA. The value of $B_{c2}$ at the lowest $T$ exceeds
the BCS Pauli paramagnetic limit~\cite{Orlando-PRB-1979}
($B_{c2}^{Pauli} = 1.8 T_s \approx 1$ T for sample \#3), which for
spin-singlet pairing is only possible in the case of strong
spin-orbit scattering. On the other hand, the absence of Pauli
limiting is expected for a triplet SC with equal-spin pairing
state~\cite{Klemm-PhysicaB-1985}.

The small ordered moment of 0.03 $\mu_{B}$ and low Curie
temperature locate UCoGe close to the FM instability (i.e. the
limit $T_C \rightarrow 0$). The proximity to the FM QCP can be
further investigated using the Ehrenfest relation for second-order
phase transitions $dT_{C}/dp = V_{m}T_{C} \Delta \alpha / \Delta
c$ (with the molar volume $V_{m} = 3.13$x$10^{-5} ~$m$^{3}$/mol).
From the estimated step-sizes in $\alpha(T)$ and $c(T)$ at $T_C$
we calculate $dT_{C}/dp = -0.25$ K/kbar. This shows that the
critical pressure $p_c$ at which magnetism vanishes is low (an
upper-bound for $p_c$ assuming a linear suppression of $T_C$ is
$\sim$ 12 kbar). In the same way we find that the SC transition
temperature ${\it increases}$ with pressure at a rate $dT_{s}/dp
\approx 0.048$ K/kbar. In the scenario of the coexistence of
$p$-state SC and FM~\cite{Fay-PRB-1980}, the increase of $T_{s}$
with pressure places UCoGe in the phase diagram on the far side of
the SC lobe with respect to the critical point (compare UGe$_{2}$
at pressures of 10-12 kbar~\cite{Saxena-Nature-2000}).
Accordingly, upon applying pressure, $T_{s}$ is predicted to pass
through a maximum before vanishing at the magnetic critical point.
The derived pressure dependencies of $T_C$ and $T_{s}$ for UCoGe
have an opposite sign compared to those for URhGe. In URhGe $T_C$
shows a monotonic increase under pressures up to 120
kbar~\cite{Hardy-PhysicaB-2005} and $T_{s}$ is suppressed with
pressure. The positive pressure dependence of $T_{s}$ in UCoGe may
explain the large difference in onset temperatures for
superconductivity in the transport and bulk properties. Positive
stress at the grain boundaries could cause a small volume fraction
of the samples to have a larger $T_{s}$.

The occurrence of SC in a FM material is naturally
explained~\cite{Fay-PRB-1980} by the formation of Cooper pairs
with parallel spin. In UCoGe the proximity to the magnetic
instability, the defect sensitivity of $T_{s}$ and the absence of
Pauli limiting are all in agreement with such a scenario. Within
the symmetry classification for orthorhombic itinerant FM
spin-triplet superconductors~\cite{Mineev-PRB-2004} the SC gap is
predicted to be anisotropic with point nodes along the magnetic
moment direction or line nodes in the plane perpendicular to the
moments. The determination of the gap function, however, requires
experiments on single crystals. In the case of URhGe, which
belongs to the same symmetry class as UCoGe, upper critical field
measurements~\cite{Hardy-PRL-2005} on a single crystal indicate a
$p$-wave polar order parameter with a maximum gap parallel to the
$a$ axis (the order moment points along the $c$
axis~\cite{Aoki-Nature-2001}). The difference of a factor 7 in the
size of the ordered moment $m_{0}$ (for URhGe the powder-averaged
moment is $m_{0} \approx 0.21 \mu _{B}$~\cite{Aoki-Nature-2001})
and the opposite pressure effects on $T_{C}$ and $T_{s}$ seem to
indicate that UCoGe and URhGe represent two different cases of
magnetically mediated SC. Indeed the recent observation of
field-induced SC~\cite{Levy-Science-2005} in URhGe was taken as
evidence for SC stimulated by a spin rotation in the neighborhood
of a quantum phase transition under high magnetic field. In the
case of UGe$_{2}$ the situation is again different as the FM to
paramagnetic transition at the critical pressure becomes first
order~\cite{Saxena-Nature-2000}. Moreover,
evidence~\cite{Pfleiderer-PRL-2002,Sandeman-PRL-2003} is available
that SC is driven by a changing Fermi surface topology associated
with a metamagnetic jump in the magnetization. Consequently,
unlike URhGe and UGe$_{2}$, UCoGe may present a genuine case of SC
at a FM quantum critical point.

In conclusion, we have demonstrated that UCoGe is a weak
ferromagnet below $T_{C} = 3$ K and becomes superconducting upon
further cooling with $T_{s} = 0.8$ K for the best sample. The
sizeable discontinuities in the thermodynamic properties at both
transition temperatures provide evidence for the bulk-like nature
of both states. The coexistence of FM and SC is unusual and
suggests SC mediated by magnetic interactions rather than by
phonons. Since both SC and FM occur at ambient pressure, UCoGe
offers a unique opportunity to elucidate the long-standing issue
of SC stimulated by critical fluctuations associated with a
magnetic quantum critical point.

This work was part of the research program of FOM (Dutch
Foundation for Fundamental Research of Matter) and COST Action P16
ECOM. Funding by the Helmholtz Association under VH-VI 127 is
gratefully acknowledged.

\end{document}